# Nonanalytic Correlation Length in Ising Systems with one Surface Defect Line

A. HUCHT
Theoretische Tieftemperaturphysik, Universität Duisburg
Lotharstraße 1, 47048 Duisburg, Germany
e-mail: fred@thp.Uni-Duisburg.de

**Abstract**

A two-dimensional Ising system with ferromagnetic coupling and one defect line at distance $L$ from the surface is solved exactly using Pfaffians. The system shows a singularity in the surface correlation length at a temperature $T_s$ which is smaller than the transition temperature $T_c$ of the bulk. Numerical studies using the transfer matrix technique suggest that this singularity is also present in an Ising system with two defect lines at distance $L$.

# Introduction

The two-dimensional Ising model with nearest-neighbour coupling is one of the best investigated models in statistical mechanics. After the famous solution obtained nearly half a century ago by Onsager[(1)] (using the transfer matrix technique) several methods have been developed to investigate the influence of restricted geometries, boundary conditions, or a line of modified bonds called a defect line[(2,3,4)]. Interface models like the solid-on-solid (SOS) model[(5)] were found to feature many of the interface effects in Ising models.

Recently, in a SOS model with two defect lines at distance $L$, a non-thermodynamic singularity in the correlation length parallel to the defect lines has been found by Upton[(6)]. The interesting question, whether this singularity also occurs in the corresponding Ising model, was already the subject of Monte Carlo simulations[(7)]. However, the existance of a singularity in the correlation length could not be shown clearly due to the finiteness of the investigated system.

In this paper a system is investigated which consists of an infinite long Ising stripe of finite width $L$ coupled with one defect line to an Ising system infinite in both directions. This system can be regarded as the special case of an infinite Ising system with two defect lines at distance $L$ where one of the defects is set equal to zero. In the exact solution of this model one finds a singularity in the surface correlation length $\xi$ at a temperature $T_S$ below the critical temperature $T_C$ of the system. With the strength of the modified bonds in the defect line tending to zero, the relation between the correlation length perpendicular to the defect line and $L$ is basically the same as in the SOS model[(6)]. For $L = 1$ the surface correlation length is expressed in a closed form and it is compared with numerical studies of Ising stripes with periodical and antiperiodical boundary conditions and two defect lines. Even for narrow stripes of width $M \leq 10$ the tendency towards a singularity in this systems is well established.

# Method

Consider a next neighbour Ising system with $N$ spins in the horizontal and $M$ spins in the vertical direction. The system has periodic boundary conditions in the $N$-direction and free boundaries in the $M$-direction. All couplings are equal to $J$, with $J > 0$, except for the couplings between the layers $L$ and $L$+1 which are denoted by $J' = \alpha J$. For $L = 1$ and $M = 5$ the system is shown schematically in Fig. 1.

The Partition Function of this system can be expressed[8] in terms of the Pfaffian of an antisymmetric $4MN$ x $4MN$ matrix **A** (with $K = J/k_B T$ and $K' = J'/k_B T$ )

$$Z = \tfrac{1}{2} 2^{NM} (\cosh K)^{N(2M-2)} (\cosh K')^N \,\mathrm{Pf}\,\mathbf{A} \qquad (1)$$

which can be evaluated[9] as a product over determinants of $2M$ x $2M$ matrices **C** as

$$(\mathrm{Pf}\,\mathbf{A})^2 = \det \mathbf{A} = \prod\nolimits_\theta c^M \det \mathbf{C} \qquad (2)$$

Here $c = \left|1 + ze^{i\theta}\right|^2$, $z = \tanh K$ and $\theta = \pi(2n-1)/N$ with $n$ running from 1 to $N$. Note that **C** is a function of $\theta$. For a system with $L = 1$ the matrix **C** is given by

$$\mathbf{C} = \begin{pmatrix} -a & b & & & & & & \\ -b & a & y & & & & & \\ & -y & -a & b & & & & \\ & & -b & a & z & & & \\ & & & -z & & & & \\ & & & & \cdots & z & & \\ & & & & -z & -a & b \\ & & & & & -b & a \end{pmatrix} \qquad (3)$$

with the matrix elements $a = c^{-1} 2iz \sin\theta$, $b = c^{-1}(1 - z^2)$ and $y = \tanh K'$. For arbitrary $L$ the element $y$ ($-y$) describing the defect line appears at position $\{2L, 2L+1\}$ ($\{2L+1, 2L\}$). The determinant of **C** can be expressed as

$$\det \mathbf{C} = c^{-M} \langle 1,0 | \mathbf{T}^{M-L} \mathbf{Y} \mathbf{T}^L | 1,0 \rangle \qquad (4)$$

with the two hermitian 2 x 2 matrices

$$\mathbf{T} = c \begin{pmatrix} b^2 - a^2 & az \\ -az & z^2 \end{pmatrix} \; ; \; \mathbf{Y} = \begin{pmatrix} 1 & 0 \\ 0 & y^2/z^2 \end{pmatrix} \qquad (5)$$

The eigenvalues and eigenvectors which satisfy $\mathbf{T}|\mathbf{t}_i\rangle = \lambda_i |\mathbf{t}_i\rangle$ are

$$\lambda_{1,2} = \tfrac{1}{2} c \left( b^2 - a^2 + z^2 \pm \sqrt{(b^2 - a^2 + z^2)^2 - 4b^2 z^2} \right) \qquad (6a)$$

$$|\mathbf{t}_i\rangle = \left| \frac{az}{a^2 - b^2 + \lambda_i}, 1 \right\rangle \qquad (6b)$$

The correlation function between two surface spins at distance $n$ is calculated as[10]

$$\langle \sigma_{1,0} \sigma_{1,n} \rangle_M = -\frac{1}{2N} \sum\nolimits_\theta e^{i\theta n} [\mathbf{C}^{-1}]_{1,1} \qquad (7)$$

for $n > 0$. The inverse of **C** is obtained from the cofactor rule and the {1,1}-element of $\mathbf{C}^{-1}$ yields

$$[\mathbf{C}^{-1}]_{1,1} = \frac{\text{cofactor}(\mathbf{C}_{1,1})}{\det \mathbf{C}} = -\frac{\langle 1,0|\mathbf{T}^{M-L}\mathbf{Y}\mathbf{T}^{L-1}\mathbf{S}|1,0\rangle}{\langle 1,0|\mathbf{T}^{M-L}\mathbf{Y}\mathbf{T}^{L}|1,0\rangle} \tag{8}$$

with the matrix

$$\mathbf{S} = -c\begin{pmatrix} a & 0 \\ z & 0 \end{pmatrix} \tag{9}$$

Up to this point the calculation is valid for arbitrary $N$ and $M$. For $N \to \infty$ the sum in eq. (7) is converted into an integral and transformed onto the unit circle in the complex plane with the substitution $\omega = e^{i\theta}$. With

$$f_{ML}(\omega) = \frac{\langle 1,0|\mathbf{T}^{M-L}\mathbf{Y}\mathbf{T}^{L-1}\mathbf{S}|1,0\rangle}{\langle 1,0|\mathbf{T}^{M-L}\mathbf{Y}\mathbf{T}^{L}|1,0\rangle} \tag{10}$$

eq. (7) reads

$$\langle \sigma_{1,0}\sigma_{1,n}\rangle_{ML} = \frac{1}{2\pi i}\oint d\omega\, \omega^{n-1} f_{ML}(\omega) \tag{11}$$

The function $f_{ML}(\omega)$ in eq. (10) has $M$ simple poles in the inner of the unit circle, all lying on the real axis and one obtains

$$\langle \sigma_{1,0}\sigma_{1,n}\rangle_{ML} = \sum_{i=1}^{M} \omega_i^{n-1}\,\text{Res}\{f_{ML}(\omega), \omega_i\} \tag{12}$$

where $\text{Res}\{f(w), z\}$ stands for the residue of the function $f$ at the point $z$. At long distances and noncritical temperatures the surface correlation function behaves as

$$\Gamma_{ML}(n) = \langle \sigma_{1,0}\sigma_{1,n}\rangle_{ML} - \langle \sigma_1\rangle_{ML}^2 \propto e^{-n/\xi_{ML}} \tag{13}$$

which can be solved for the inverse correlation length $\xi_{ML}^{-1}$ to give

$$\xi_{ML}^{-1} = \lim_{n\to\infty} \frac{\partial}{\partial n} \ln \Gamma_{ML}(n) \tag{14}$$

For finite $M$ the term $\langle \sigma_1\rangle_{ML}$ in eq. (13) is equal to zero and the correlation length yields

$$\xi_{ML}^{-1} = \ln \omega_1 \tag{15}$$

where $\omega_1$ denotes the pole of $f_{ML}(\omega)$ closest to 1.
In the limit $M \to \infty$ the calculation will be restricted to the case $T < T_C$. Now eq. (10) reduces to

$$f_L(\omega) = \lim_{M\to\infty} f_{ML}(\omega) = \frac{\langle \mathbf{t}_1|\mathbf{Y}\mathbf{T}^{L-1}\mathbf{S}|1,0\rangle}{\langle \mathbf{t}_1|\mathbf{Y}\mathbf{T}^{L}|1,0\rangle} \tag{16}$$

where $|\mathbf{t}_1\rangle$ from eq. (6) is the eigenvector of **T** with the eigenvalue larger in magnitude. Below $T_C$ the correlation length in eq. (15) diverges because $\lim_{M\to\infty} \omega_1 = 1$ and a spontanous surface magnetisation occurs which can be expressed as

$$\langle\sigma_1\rangle_L^2 = \frac{1}{2}\mathrm{Res}\{f_L(\omega),1\} \qquad (17)$$

The correlation function $\Gamma_L(n)$ now reads

$$\Gamma_L(n) = \frac{1}{2\pi i}\oint_C d\omega\, \omega^{n-1} f_L(\omega) \qquad (18)$$

The contour of integration $C$ in eq. (18) can be contracted but must enclose all singularities of $f_L(\omega)$ in the inner of the unit circle. With increasing $n$ the long distance behaviour of $\Gamma_L(n)$ is dominated by the singularity of $f_L(\omega)$ closest to 1 which will be denoted $\omega_2$, and with eq. (14) the surface correlation length is given by

$$\xi_L^{-1} = \ln\omega_2 \qquad (19)$$

This correlation length describes the decay of the correlation function $\langle\sigma_{1,0}\sigma_{1,n}\rangle_L$ onto its limiting value $\langle\sigma_1\rangle_L^2$ and will be investigated in the next section.

# Exact Results

In general finding the explicit representation of $\omega_2$ in eq. (19) involves the solution of a polynomial equation of the order $2L$. For $L = 1$ the correlation length is expressed in a closed form. One finds

$$f_1(\omega) = \frac{2(\omega^2-1)^2 z^4 - y^2(z+\omega)(\omega z+1)\left(\sqrt{R(\omega)} + (z^2+1)(\omega^2 z + \omega z^2 + z - \omega)\right)}{(\omega^2-1)z\left[2z^2(z-\omega)(\omega z-1) + y^2\left(\sqrt{R(\omega)} + (z^2+1)(\omega^2 z + \omega z^2 + z - \omega)\right)\right]} \qquad (20)$$

with

$$R(\omega) = (z^2+\omega z - z + \omega)(z^2 + \omega^{-1}z - z + \omega^{-1})(z^2 + \omega z + z - \omega)(z^2 + \omega^{-1}z + z - \omega^{-1}) \qquad (21)$$

The analytic structure of $f_1(\omega)$ is shown in Fig. 2. $\alpha_i$ and $\alpha_i^{-1}$ represent the zeroes of $R(\omega)$. These square root singularities are mutually connected with branch cuts. It should be noted that $\alpha_i$ does not depend on $y$ for $y \neq 0$. For $\omega_2$ there exist two different cases: Below a certain temperature $T_S$ one finds an isolated pole $\gamma > \alpha_2$ as shown in Fig. 2, so $\xi^{-1} = \ln\gamma$ with the correlation length fulfilling

$$2\,\mathrm{ch}\,\xi^{-1} = \mathrm{cth}\,2K\left\{(\mathrm{ch}\,2K'+1) + (\mathrm{ch}\,2K'-1)\sqrt{1-\frac{\mathrm{ch}\,2K+1}{\mathrm{ch}^2\,2K(\mathrm{ch}\,2K-\mathrm{ch}\,2K')}}\right\} \tag{22a}$$

At $T_S$ $\gamma$ is equal to $\alpha_2$. Above $T_S$ the pole $\gamma$ vanishes and with $\xi^{-1} = \ln\alpha_2$ one obtains the well known result for the surface correlation length below $T_C$[11]

$$\xi^{-1} = 2K + \ln\tanh K \tag{22b}$$

It is interesting to note that only in the limit $\alpha \to 0$ there is a kink in $\xi^{-1}(T)$ while $\partial\xi^{-1}/\partial T$ always shows a kink. The resulting inverse correlation length is shown in Fig. 3.
The temperature $T_S$ can be calculated for arbitrary $L$. Setting $\gamma = \alpha_2$ one obtains an exact relation between $K$, $K'$ and $L$ at the singular point $T_S$ :

$$\tanh^2 K_s' = \frac{(\cosh 2K_s - 1)\left(\tfrac{1}{2}\sinh 4K_s - \cosh 2K_s - L^{-1}\right)}{\cosh 2K_s(\cosh 2K_s + 1)(\sinh 2K_s - 1)} \tag{23}$$

For $J$ = 1 and several values of $L$ the resulting temperature $T_S$ vs. the defect line coupling $J'$ is depicted in Fig. 4. Clearly $T_S$ tends to zero for all $L$ when $J' \to J$. In the limit $\alpha \to 0$ $T_S$ can also be regarded as the temperature at which an Ising stripe of width $L$ has the same correlation length as an infinite system on its surface. In this limit eq. (23) will be investigated in further detail and is solved for $L^{-1}$ to give

$$L^{-1} = \tfrac{1}{2}\sinh 4K_s - \cosh 2K_s \tag{24}$$

This function is a forth order polynomial which can be solved to give the value of $K_s(L^{-1})$ at which the correlation length is singular. This expression is inserted into eq. (22b) and expanded into a series in $L^{-1}$ around 0. The result is

$$\xi(L) = L + \frac{1}{\sqrt{2}} - \frac{1}{12}L^{-1} + \frac{1}{48\sqrt{2}}L^{-2} + \frac{11}{720}L^{-3} - \frac{109}{2560\sqrt{2}}L^{-4} + O\left(L^{-5}\right) \tag{25}$$

Thus the value of $\xi$ at the singular point is approximately $L + \sqrt{1/2}$ for large L, while for $\alpha \neq 0$ $\xi$ remains proportional to $L$, but has a modified slope (a calculation yields $\xi/L \approx 1 - 1.13\,\alpha^2 + O(\alpha^4)$ for large L). Note that the theory of finite size scaling[12] already predicts that $\xi \propto L$, because $T_S$ converges against $T_C$ with increasing $L$.
With eq. (25) this system shows the same behaviour as in the SOS model considered in reference (6), where the result was $2\xi_\perp \approx L$ at the singular point with $\xi_\perp$ denoting the correlation length perpendicular to the two defect lines in the SOS model. In the Ising model considered here one obtains the same result; the correlation length perpendicular to the surface is simply the bulk correlation length $\xi_B = \xi/2$ which is unaltered by the defect line.

Finally we calculate the spontaneous magnetisation in the first row which is obtained from eq. (17). For simplicity only the result for $L = 1$ is given here

$$\langle\sigma_1\rangle_1^2 = \frac{e^{4K}\coth 2K(\sinh 2K - 1)}{\coth^2 K'(\cosh 2K - 1) + 2e^{2K}\cosh 2K(\sinh 2K - 1)} \tag{26}$$

which clearly is an analytic function for all temperatures below $T_C$ when $\alpha > 0$. This result and numerical evaluation of the specific heat and the surface susceptibility near $T_S$ suggest that the singularity is non-thermodynamic. In Ref. (6) it is shown that the free energy in the SOS model is indeed analytic at $T_S$.

# Numerical Results

The exact results presented in this paper are for the limiting case of an Ising system with two defect lines where one of the defect lines is set equal to zero. It is an interesting question how the singularity in the correlation length parallel to the defect line depends on boundary conditions. To test the general behavior the model is compared with an Ising system which is periodic in both directions having two defects of strength $\alpha$ and $\beta$, respectively. The calculation has been done numerically for Ising stripes of infinite length and width $M$, $M \leq 10$, with the usual transfer matrix (TM) method. The distance $L$ between the defect lines is set to $L = 1$. The inverse correlation lengths of a system of width $M$ are calculated numerically exact through the eigenvalues $\lambda_1 > \lambda_2 > \lambda_3 > \dots$ of the TM as[13]

$$\xi_j^{-1} = \ln\frac{\lambda_1}{\lambda_j} \tag{27}$$

While $\xi_2^{-1}$ vanishes with increasing $M$ below $T_C$, $\xi_3^{-1}$ converges against the bulk value at these temperatures. The results for three different combinations of defect lines are presented. They are compared to the well known correlation length of the one dimensional Ising system in an external field

$$\xi_{1D}^{-1} = \ln\frac{\cosh h + \sqrt{e^{-4K} + \sinh^2 h}}{\cosh h - \sqrt{e^{-4K} + \sinh^2 h}} \tag{28}$$

with $h = H/k_B T$. The field $H$ is set to the mean field acting on the spins in the chain, neglecting the interaction across the two defect lines, namely $H = (\alpha + \beta)\langle\sigma_1\rangle$ with the surface magnetisation $\langle\sigma_1\rangle$ of the two dimensional Ising model[14]. $\langle\sigma_1\rangle$ can also be derived from eq. (26) for $K' \to K$. For $\alpha = 0.5$ and $\beta = 0$ (Fig. 5) the results clearly show the tendency towards the exact solution. The general behaviour of the results, especially their deviation from eq. (28) and convergency, remains the same for the periodic ($\alpha = \beta = 0.5$, Fig. 6) and for the antiperiodic ($\alpha = -\beta = 0.5$, Fig. 7) case. These results strongly suggest that the singularity in the correlation length is also present in an Ising system with two defect lines at distance *L*.

# Conclusions

In this paper it was shown that in a system consisting of two coupled subsystems with different dimensionalities $D_1$ = 1 and $D_2$ = 2 the correlation length parallel to the boundary surface shows a singularity at a temperature between the critical temperatures of the two uncoupled systems. This singularity is proposed to be present also in higher dimensions, for example when a two dimensional system is coupled to a three dimensional bulk. A simple argument for the presence of this singularity can be given at least in the limit $\alpha \to 0$: Consider the transfer matrix (TM) **U** of a system of arbitrary dimensionality that consists of two subsystems infinite in at least one dimension and coupled with a defect "plane" of strength $\alpha$. Let system 1 have the higher critical temperature. For $\alpha \to 0$ the TM reduces to an outer product $\mathbf{U} = \mathbf{U}^{(1)} \otimes \mathbf{U}^{(2)}$ with $\mathbf{U}^{(1)}$ and $\mathbf{U}^{(2)}$ being the TM of system 1 and 2, respectively. The eigenvalues $\lambda_i$ of **U** are all possible products of the eigenvalues $\mu_1 > \mu_2 > \cdots$ of $\mathbf{U}^{(1)}$ and $\nu_1 > \nu_2 > \cdots$ of $\mathbf{U}^{(2)}$ which may be degenerated. Sorting the eigenvalues of **U** by magnitude one obtaines

$$\lambda_1 = \mu_1 \nu_1 \;, \quad \lambda_2 = \max(\mu_2 \nu_1, \mu_1 \nu_2) \tag{29}$$

So $\lambda_2$ is nonanalytic when it is passing from $\mu_2 \nu_1$ to $\mu_1 \nu_2$. This indeed occurs in the temperature range $T_c^{(2)} < T < T_c^{(1)}$ when $\mu_1 / \mu_2 = \nu_1 / \nu_2$ or $\xi^{(1)} = \xi^{(2)}$, because $\xi^{(1)}$ has a pole at $T_c^{(1)}$ and shrinks with increasing temperature, while $\xi^{(2)}$ grows and has a pole at $T_c^{(2)}$.

The singularity in the surface correlation length found in this paper is believed to be non-thermodynamic because the magnetisation is an analytic function of temperature and numerical studies of the free energy and other thermodynamic quantities as the specific heat or the surface susceptibility show no influence in the corresponding temperature range. In the limit $\alpha \to 0$ eq. (29) shows that the free energy is an analytic function of temperature at $T_S$.

Another result of the calculation, namely eq. (25), gives the amplitude at which the correlation length of a stripe of width *L* diverges when $T \to T_c^-$ of the infinite system and $L(T) \to \infty$. While it is well known that at fixed temperature $T = T_c$ for free boundary conditions and large *L* one obtaines[15]

$$\xi^{-1}(L^{-1}, T) \approx \frac{\pi}{2} L^{-1} \tag{30}$$

Eq. (25) states that for $T \to T_c^-$ and $L(T)$ given by eq. (24) the result is

$$\xi^{-1}\left(L^{-1},T\right) \approx L^{-1} \tag{31}$$

This result may be of some interest for the theory of conformal invariances[15].

# Acknowledgements

I would like to thank Prof. Dr. K. Usadel and Dipl. Phys. M. Schröter for helpful discussions and critical reading of the manuscript. Furthermore I would like to thank W. Selke for sending me his preprints prior to publication. This work was supported by the Deutsche Forschungsgemeinschaft through SFB 166.

# References

[1] L. Onsager, *Phys. Rev.* **65**, 117 (1944).

[2] D. B. Abraham, in *Phase Transitions and Critical Phenomena*, Vol. 10, Academic, New York (1986).

[3] M. E. Fisher and A. E. Ferdinand, *Phys. Rev. Lett.* **19**, 169 (1967).

[4] D. B. Abraham, *Phys. Rev. Lett.* **44**, 1165 (1980).

[5] H. N. V. Temperley, *Proc. Camb. Phil. Soc.* **48**, 683 (1952).

[6] P. J. Upton, Preprint (1991).

[7] W. Selke, Preprint (1991); W. Selke, N. M. Švrakić and P. J. Upton, Preprint (1991).

[8] P. W. Kasteleyn, *J. Math. Phys.* **4**, 287 (1963).

[9] B. McCoy and T.T. Wu, *The Two-Dimensional Ising Model*, Harvard University Press, Cambridge (1973).

[10] Ref. 10, Chapter VII, B. M. McCoy and T. T. Wu, *Phys. Rev.* **162**, 436 (1967).

[11] See for example Ref. 10.

[12] M. N. Barber, in *Phase Transitions and Critical Phenomena*, Vol. 8, Academic, London (1983).

[13] M. E. Fisher and V. Privman, *Phys. Rev.* **B32**, 447 (1985).

[14] Ref. 10, Chapter VI.

[15] J. L. Cardy, *J. Phys. A: Math. Gen.* **17**, L385-L387 (1984).

# Figure Captions

Figure 1: Schematic structure of the system for $L = 1$ and $M = 5$.

Figure 2: Analytic structure of $f_1(\omega)$ from eq. (20) at a temperature below $T_S$.

Figure 3: Reduced inverse surface correlation length, $T\xi^{-1}$, vs. temperature for $J = 1$ and various values of $\alpha$. Full lines are for $T$ below $T_S$ (eq. 22a), dotted line for $T$ above $T_S$ (eq. 22b).

Figure 4: Temperature $T_S$ vs. defect line coupling $J'$ as given in eq. (22) for $J = 1$ and several values of $L$.

Figure 5: Comparison between the exact solution (full and dotted line) and numerical results for $J = 1$, $L = 1$ and different $M$ (dashed lines). The broken line represents the solution of eq. (28).

Figure 6: Numerical results for a system with $J = 1$, two defect lines $\alpha = \beta = 0.5$ at distance $L = 1$ and different $M$ (dashed lines). The broken line represents the solution of eq. (28).

Figure 7: Numerical results for a system with $J = 1$, two defect lines $\alpha = -\beta = 0.5$ at distance $L = 1$ and different $M$ (dashed lines). The broken line represents the solution of eq. (28).

Figure 1

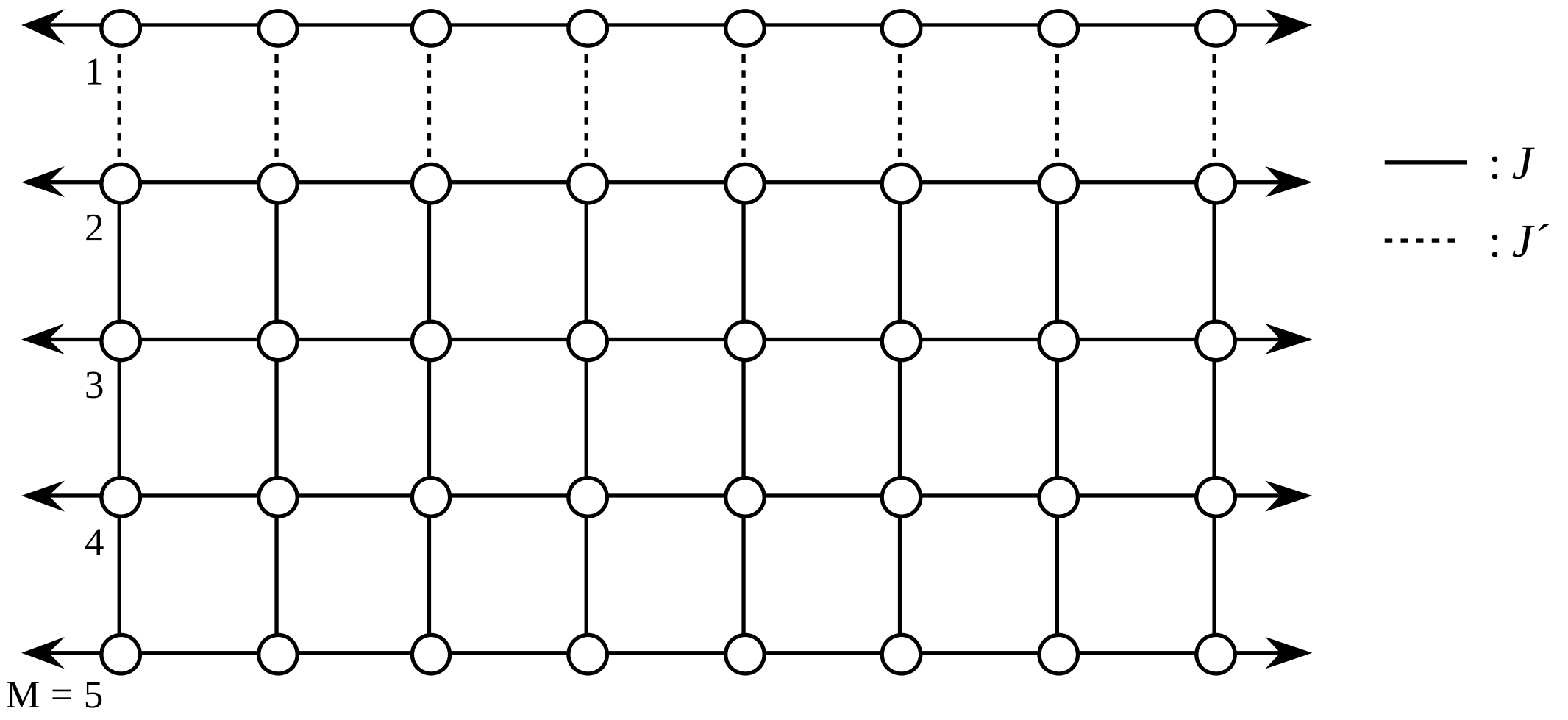

Figure 2

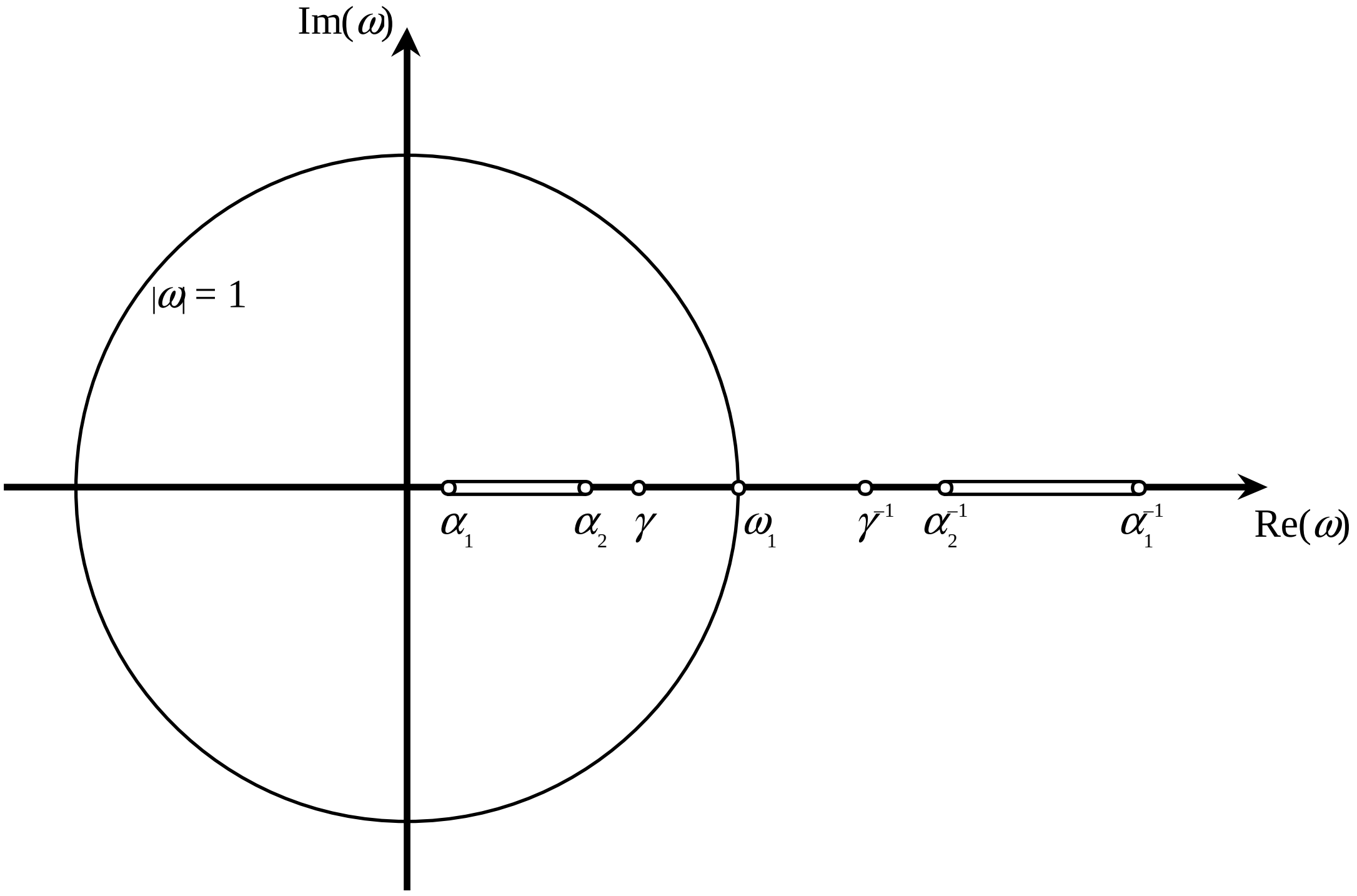

Figure 3

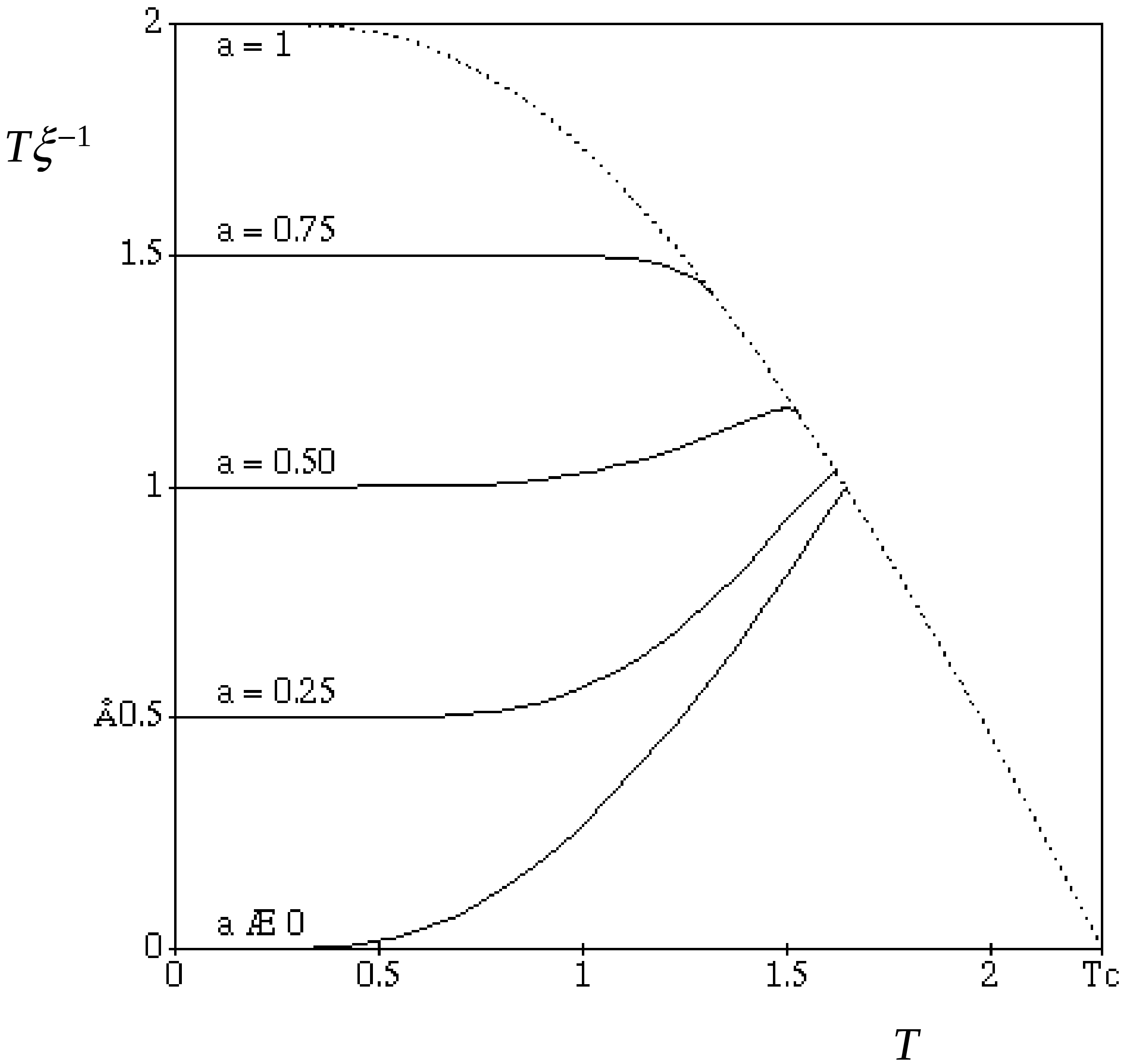

Figure 4

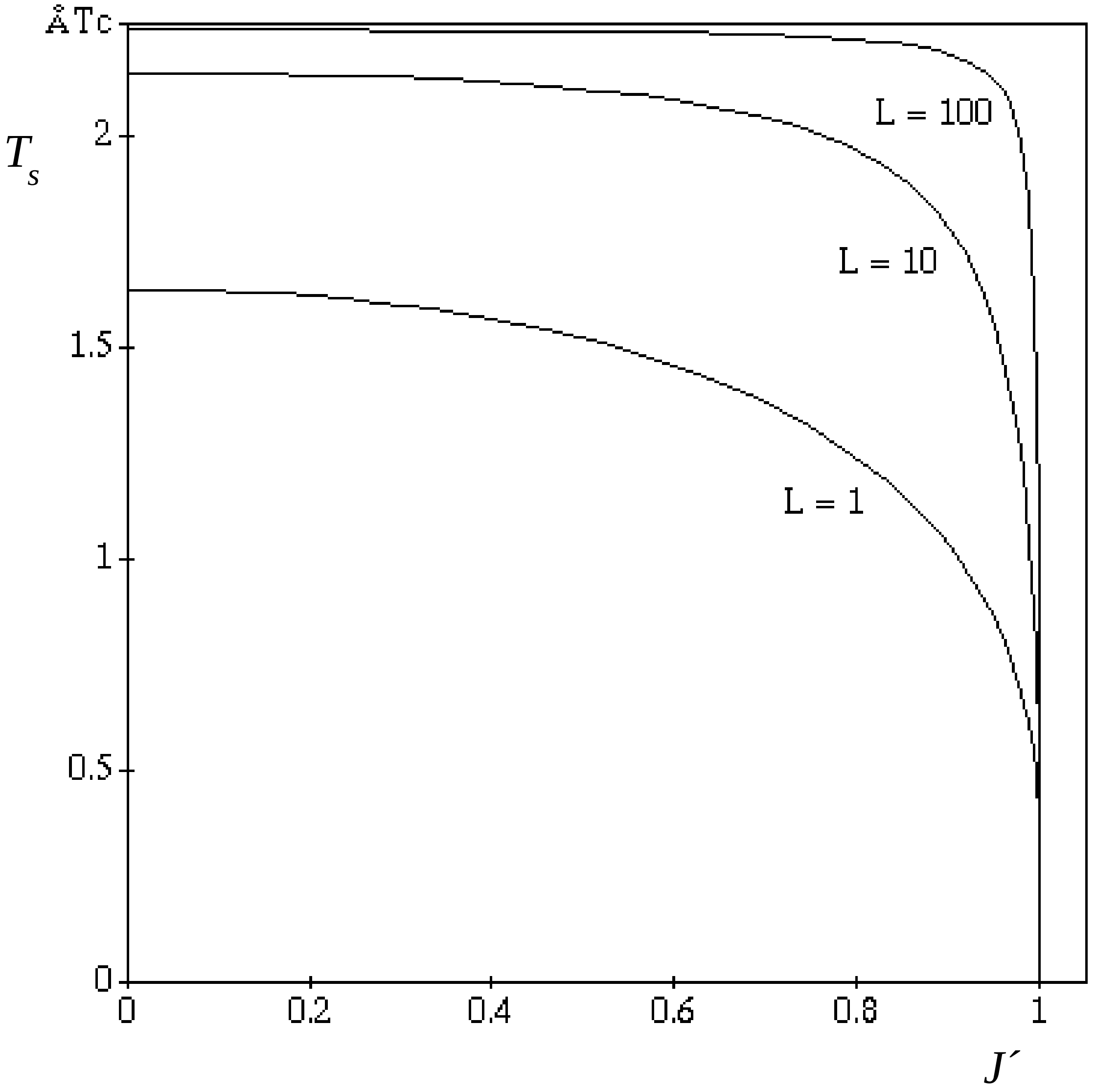

Figure 5

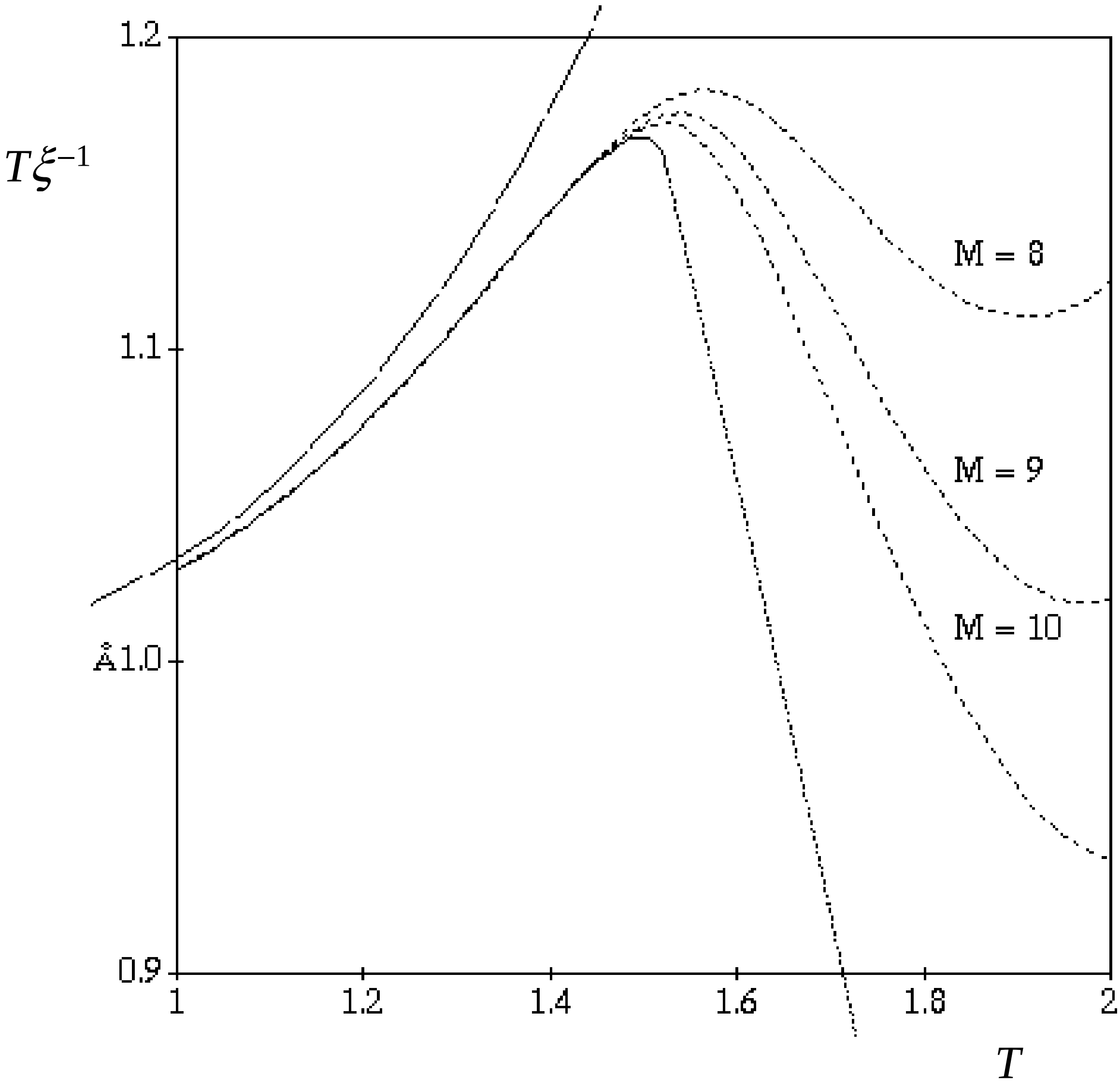

Figure 6

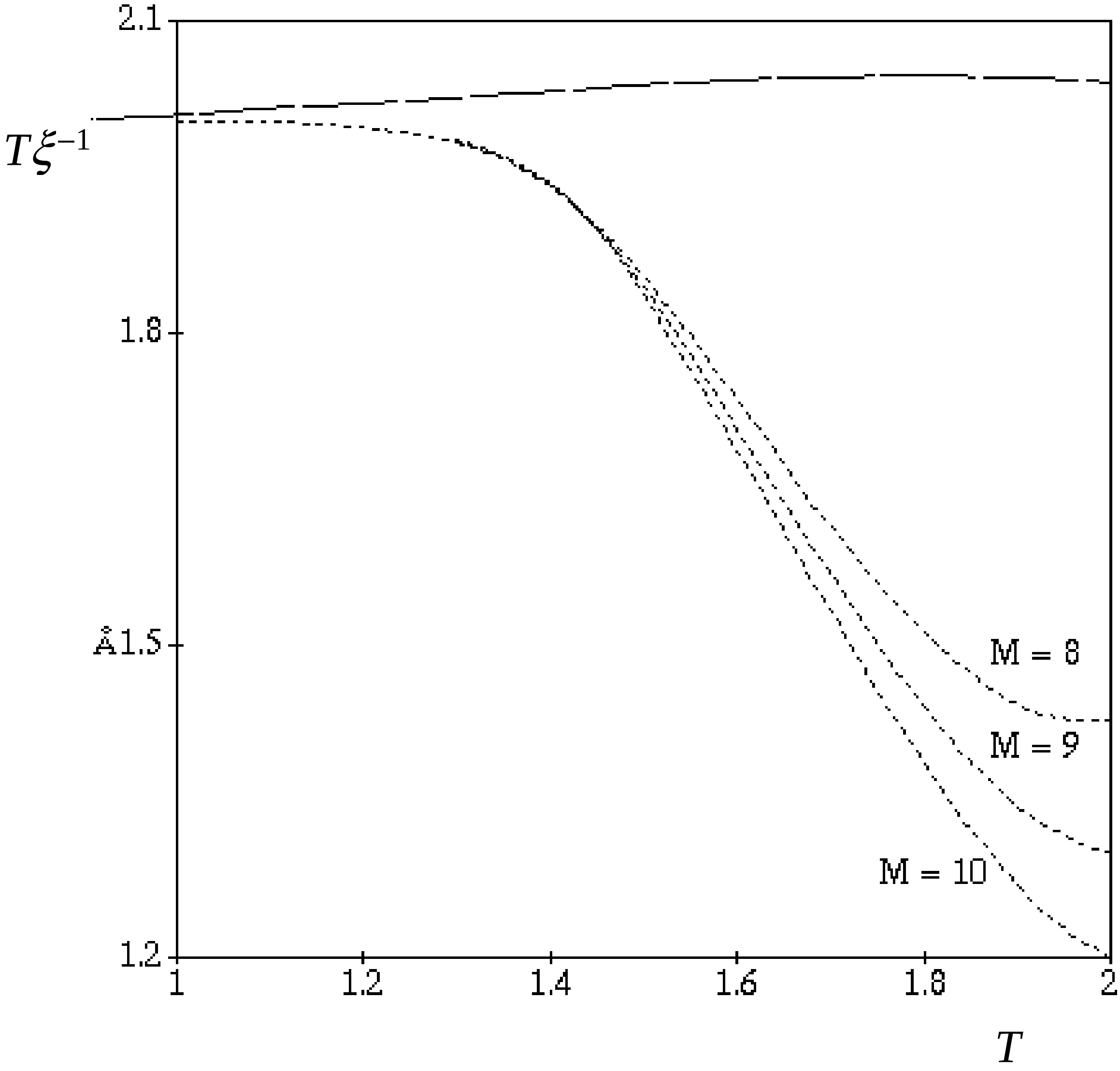

Figure 7

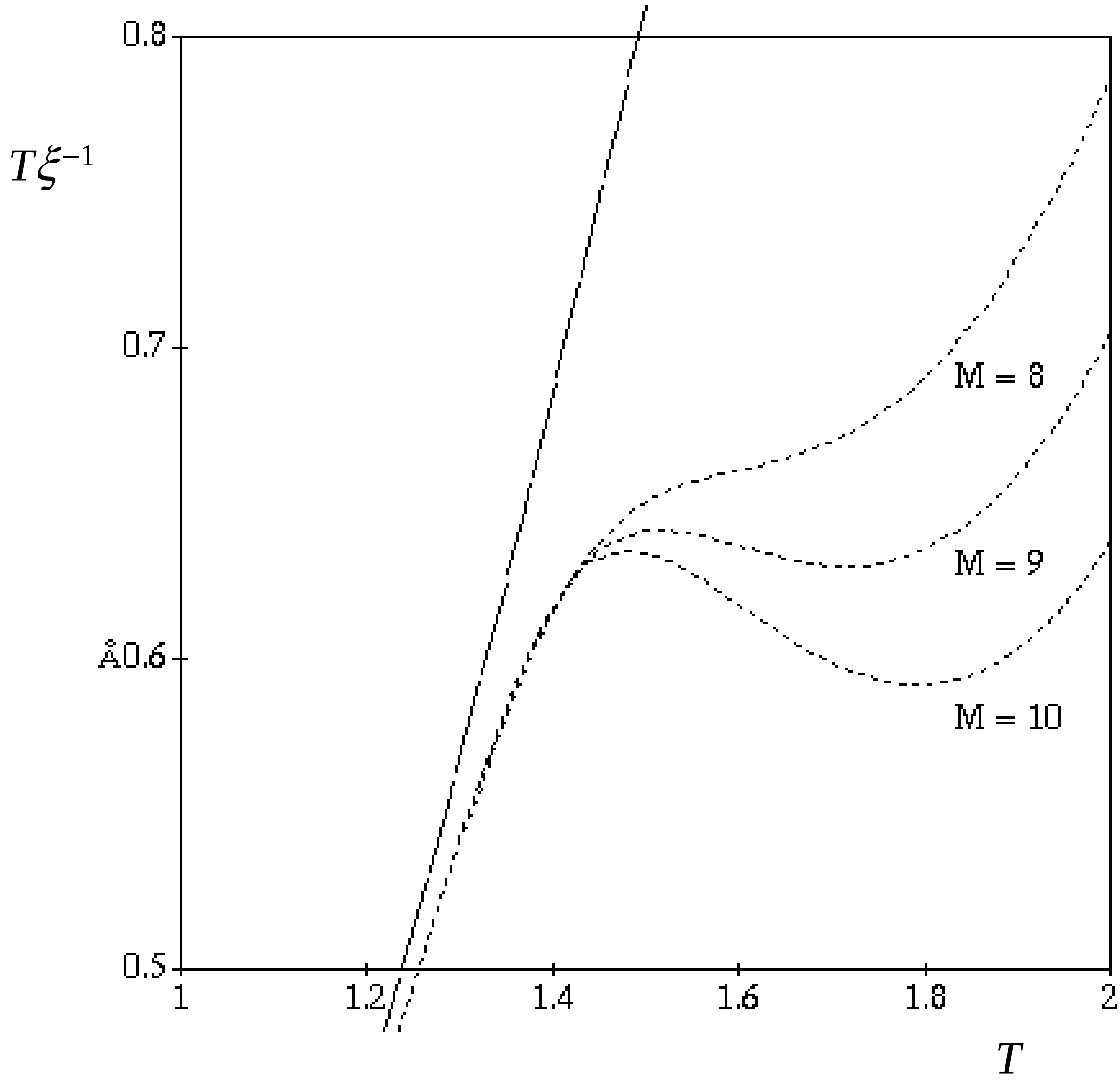